\documentclass[12pt]{iopart}
\usepackage{graphicx}% Include figure files
\usepackage{dcolumn}% Align table columns on decimal point
\usepackage{bm}% bold math
\usepackage{cite}
\usepackage{fancyhdr}
\begin{document}

\title{Spin dynamics at the singlet-triplet crossings in double quantum dot}

\author{J.~S\"arkk\"a and A.~Harju}

\address{Helsinki Institute of Physics and Department of Applied Physics, Aalto University, 
FI-02150 Espoo, Finland}

\ead{jani.sarkka(at)tkk.fi}

\begin{abstract}
We simulate the control of the
spin states in a two-electron double quantum dot
when an external detuning potential is used for
passing the system through an anticrossing.
The hyperfine coupling of the
electron spins with the surrounding nuclei causes
the anticrossing of the spin states but also the
decoherence of the spin states.
We calculate numerically the singlet-triplet decoherence
for different detuning values and find a good agreement
with experimental measurement results of a similar setup.
We predict an interference effect due to the coupling of
$T_0$ and $T_+$ states.

\end{abstract}


\tableofcontents

\section{Introduction}

The development of a working quantum computer, based on
electron spins confined in quantum dots \cite{loss98,hanson:1217},
has been the objective of a large field of experimentalists
and theorists. One of the most important steps in realizing
a spin qubit is achieving rapid control of the spin states.
For two-electron quantum dots, a method based on electrical
control of a singlet-triplet qubit was presented by Petta et al. \cite{petta05},
enabling control times of a nanosecond time scale. Recently,
the control of a singlet-triplet qubit based on dynamic nuclear polarization and
inhomogeneous magnetic field has been demonstrated \cite{foletti:903}.

The electrical control of spin states is based on Landau-Zener
transitions between singlet and triplet states. These transitions take place when external 
detuning voltage is swept over singlet-triplet crossings. 
The coupling between singlet and triplet states in GaAs is due to
the hyperfine interaction between spins of the confined electrons and spins of the
surrounding nuclei \cite{khaetskii:186802,merkulov:205309}. The transition probability depends on the
sweeping rate of
the detuning voltage and on the strength of the hyperfine coupling according to the
Landau-Zener formula \cite{landau32,zener32,wittig:8428,lzs10}. The external magnetic field
splits the triplet states into three states separated by the Zeeman splitting, 
$T_+$, $T_0$ and $T_-$. 
For certain detuning values, $S-T_+$ and $S-T_0$ energy differences are smaller than the
hyperfine coupling. Hence, transitions from the singlet to triplet states are possible and result
in a nonvanishing occupation probability of the triplet states.

Petta et al.\cite{petta10,burkard10} studied experimentally consecutive Landau-Zener transitions 
between $S$ and $T_+$ states.
They initialized a two-electron double dot so that the two electrons are in a single
potential well. The Pauli exclusion principle determines that the electrons are in a singlet
state. After the initialization, they rapidly moved the detuning to a value $\varepsilon$, 
where the potential has a double-well form and the electrons can tunnel between the wells.
They held the detuning fixed for some time, and then rapidly moved 
the detuning back to its starting value, where they measured the 
probability to return as a singlet. They observed an interference 
pattern of decaying oscillations of the singlet probability as a 
function of $\varepsilon$ and time.
The decay of the oscillations can be explained by the fluctuations of the nuclear spins. 
This experiment
has recently been studied theoretically in Ref.~\cite{ribeiro:115445}.

In this paper, we study the electrical control of a two-electron double quantum dot using 
Landau-Zener transitions, following the cycle given above. We calculate numerically the dynamics 
of the setup
using a 4 $\times$ 4 Hamiltonian matrix,
derived by Coish and Loss \cite{coish:125337}, and evaluate numerically 
the probabilities of the singlet and triplet states as a function of time.
We observe a similar interference pattern as observed in the experiments \cite{petta10}.
The occupation of the $T_0$ state has to be taken into account for smaller detunings.
The combination of $S-T_0$ and $S-T_+$ oscillations creates a secondary interference pattern,
which demands higher precision in order to be observed in the measurements.

\section{Model and Method}
\begin{figure}
\mbox{}\hfill
\includegraphics[width=0.45\columnwidth]{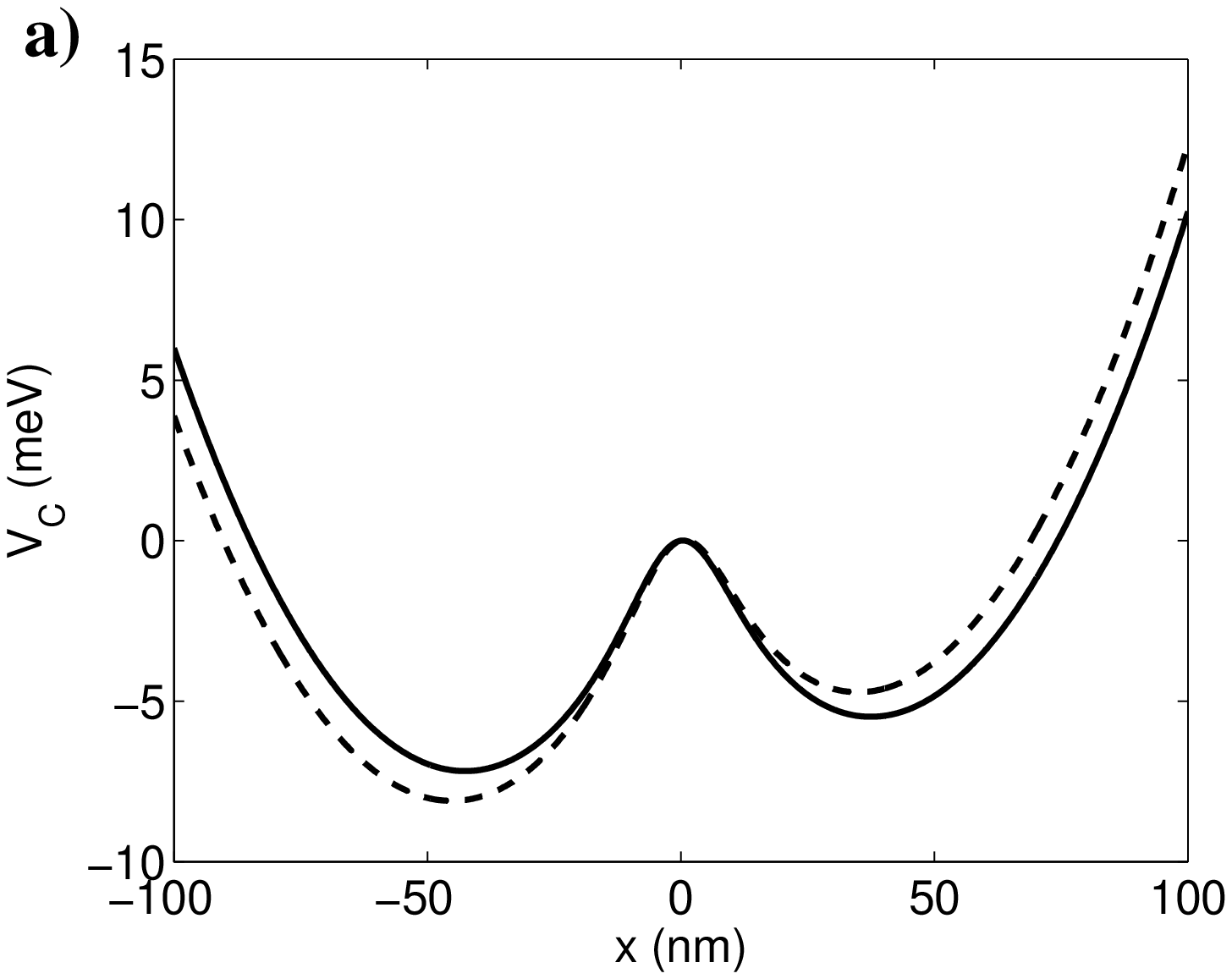}
\includegraphics[width=0.45\columnwidth]{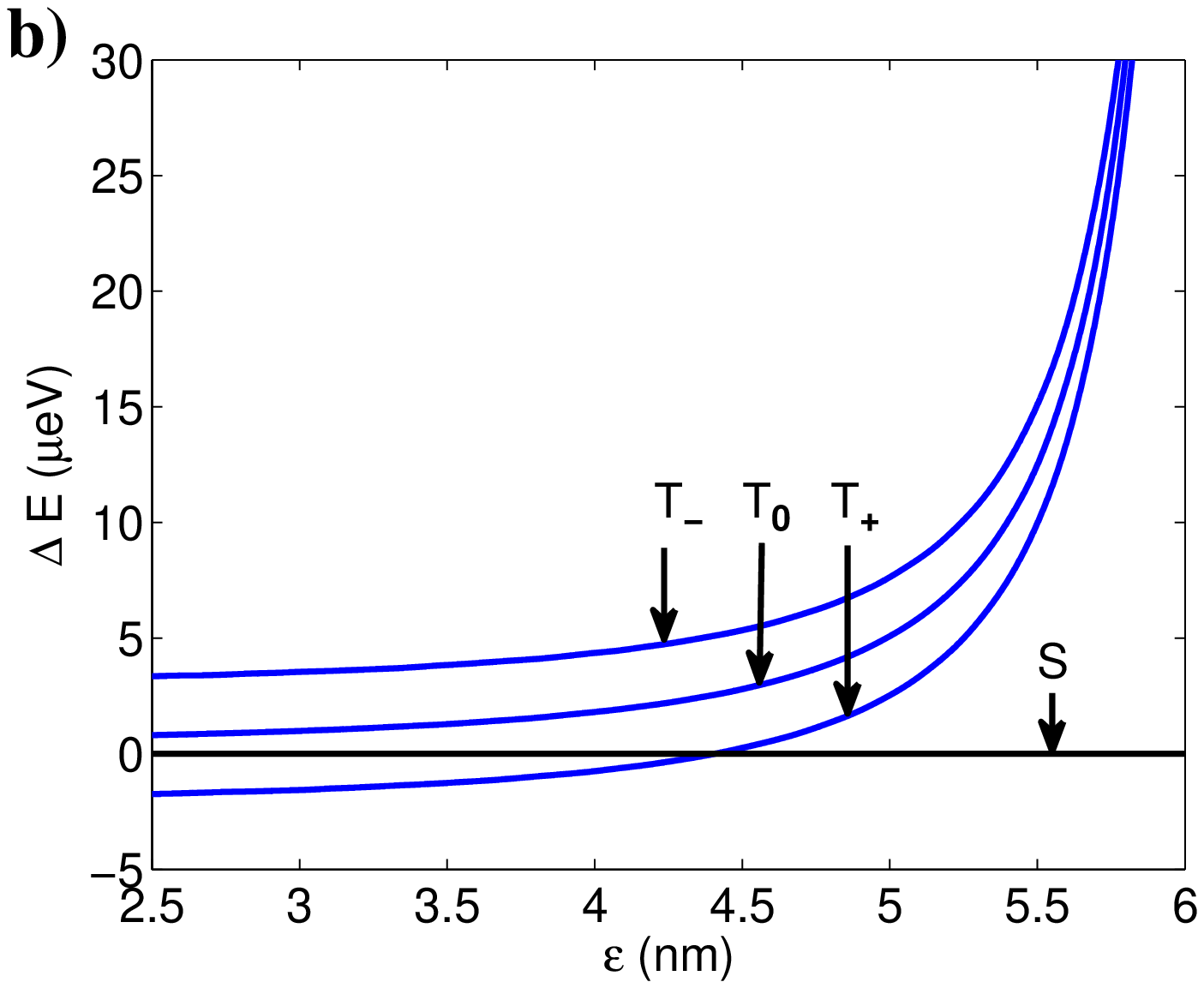}
\caption{(Color online)
\label{fig:triplets}
(a) The confinement potential $V_C$ with $L$=40 nm, $\varepsilon$=5.3 nm (dashed), and 
$\varepsilon$=2.7 nm (solid).
(b)
The energy difference $\Delta E$ between the singlet state
$S$ and triplet states $T_-,T_0$ and $T_+$ as a function
of the detuning $\varepsilon$. In the calculation of
these energies, the hyperfine field is set to zero.
The difference of $S$ and $T_0$ states is the exchange energy $J_0$.
$T_-$ and $T_+$ states are separated from the $T_0$ state by the
Zeeman splitting $\epsilon_Z$. 
The dot distance is 80 nm and magnetic field $B$=100 mT. 
$S$ and $T_+$ states cross at $\varepsilon$=4.4 nm.}
\end{figure}

We model the two-electron system with the Hamiltonian
\begin{equation}
H=\sum_{i=1}^{2} \Bigg(\frac{\Big(-i\hbar \nabla_{i}
-\frac{e}{c}\mathbf{A}\Big)^{2}}{2m^{*}}
+V(\mathbf{r}_{i},\mathbf{s}_{i}) \Bigg)
+\frac{e^{2}}{\epsilon r_{12}},
\end{equation}
where the effective mass $m^*$=$0.067m_e$ 
and permeability $\epsilon$=12.7
are material parameters for GaAs.
The external potential $V$ is divided
into two parts, confinement potential 
$V_C$ and potential due to the Zeeman interaction $V_Z$.
The smooth two-minima confinement potential, shown in Fig.~\ref{fig:triplets}(a),
is of the form
\begin{equation}
\label{eq:pot}
V_C(\mathbf{r})=\frac{1}{2}m^{*}\omega^2[x^{2}-\lambda\tanh(x/\delta)x-\varepsilon x]
+\frac{1}{2}m^{*}\omega^2 y^{2},
\end{equation}
where the confinement strength is $\hbar \omega$=3.0 meV and
the parameter $\delta$=10 nm causes smoothing of the
potential around $x$=0. The detuning voltage is taken into
account by the parameter $\varepsilon$.
The $\lambda$ 
parameter is used to fix the coordinates of the potential
minima to $x=\pm L$, $y=0$, which is obtained using
\begin{equation}
\lambda=L\delta/[L(1/\cosh^{2}(L/\delta))+\delta\tanh(L/\delta)].
\end{equation}
We use $L$=40 nm, and the distance of the dots is $2L$=80 nm.
The potential caused by the Zeeman interaction is
\begin{equation}
V_{Z}(\mathbf{r},\mathbf{s})=g^{*}\mu_{B}\mathbf{B}(\mathbf{r})\cdot
\mathbf{s},
\end{equation}
where the Land\'e factor of GaAs is $g^{*}$=-0.44.
The magnetic field can be divided into a homogeneous 
external magnetic field $\mathbf{B}_{ext}$ and an inhomogeneous random
hyperfine field $\mathbf{B}_{nuc}(\mathbf{r})$.

We discretize the Hamiltonian using finite difference method 
and determine its eigenvalues using Lanczos diagonalization 
\cite{sarkka:245315}.
We set a numerical grid on the quantum dot, 30 grid points
in x-direction and 15 points in y-direction.
We simulate the fluctuations of the hyperfine field,
causing the decay of the singlet-triplet oscillations,
by constructing a random inhomogeneous hyperfine field. 
We evaluate the  fluctuations of the hyperfine field
by assigning a random hyperfine field vector to each grid point.

In Fig.~\ref{fig:triplets}(b), 
the energy differences between the eigenenergies of the singlet and triplet 
states are shown as a function of detuning for $B$= 100 mT.
The initial value of the detuning is chosen higher than
5.5 nm so that the system is initialized in the singlet state.
When the detuning is lowered, the energies of the triplet
states approach the energy of the singlet state. Around $\varepsilon$= 4.4 nm, 
$S$ and $T_+$ states become degenerate. 
For lower detunings, the energy of the $T_0$
state approaches asymptotically the singlet energy.
For detunings lower than $\varepsilon$ = 5 nm, the singlet-triplet energy
differences are so small that transitions from singlet
to triplet states are possible. The coupling of the different spin states
is via the hyperfine field term.
For $\varepsilon<$ 3.3 nm, the
$T_0$ state is closer in energy to $S$ state 
than $T_+$ state. Thus, the $T_0$ state gains
larger probability than $T_+$ state in the transition
from the $S$ state. We observe that the energy differences are not
linear as a function of the detuning. A linear approximation
would be valid very near to the crossing points.
Hence, it is necessary to use energy values calculated here.

The energies of the excited states above the four lowest-lying 
states are much higher and need not to be taken into account
in our analysis.
We approximate the dynamics of the system using a
4 $\times$ 4 Hamiltonian, constructed in the basis of 
the singlet and three triplet states.
For the integrals needed in the calculation of the elements of the Hamiltonian
matrix, we use the shorthand notation
\begin{equation}
h_{i}^{\alpha}=
 g^*\mu_B 
\int \int
d\mathbf{r}_{1}
d\mathbf{r}_{2}
\psi_{S/T}^{*}(\mathbf{r}_{1},\mathbf{r}_{2})
 B^{\alpha}(\mathbf{r}_{i})
\psi_{S/T}(\mathbf{r}_{1},\mathbf{r}_{2}),
\end{equation}
where the space parts of the wave functions  
$\psi_{S/T}^{*}, \psi_{S/T}$ are either the singlet
or triplet wave functions, depending on
the calculated matrix element, 
$i$ is the index of the confined electron (1 or 2),
and $\alpha$ is $x$,~$y$, or $z$.
For brevity, we introduce variables \cite{coish:125337}
that depend on $h_{1,2}$
\begin{equation}
h^{\alpha}=\frac{1}{2}(h_{1}^{\alpha}+h_{2}^{\alpha}),\
\delta h^{\alpha}=\frac{1}{2}(h_{1}^{\alpha}-h_{2}^{\alpha}),
\end{equation}
\begin{equation}
h^{\pm}=h^{x}\pm ih^{y},\
\delta h^{\pm}=\delta h^{x}\pm i \delta h^{y}.
\end{equation}
With these variables, the expression for the Hamiltonian matrix $H_{eff}$ 
in the basis \{$S$,~$T_+$,~$T_0$,~$T_-$\} reads \cite{coish:125337}
\begin{equation}
%\begin{displaymath}
H_{eff}(\varepsilon)=
\left(\begin{array}{cccc}
0 & -\delta h^{+}/\sqrt{2} & \delta h^{z} & \delta h^{-}/\sqrt{2} \\
-\delta h^{-}\sqrt{2} & J_0(\varepsilon)+\epsilon_{Z}+h^{z} & h^{-}/\sqrt{2} & 0 \\
\delta h^{z} & h^{+}/\sqrt{2} & J_0(\varepsilon) & h^{-}/\sqrt{2} \\
\delta h^{+}/\sqrt{2} & 0 & h^{+}/\sqrt{2} & J_0(\varepsilon)-\epsilon_{Z}-h^{z}
\end{array} \right),
%\end{displaymath}
\end{equation}
where $\epsilon_{Z}$=$g^{*}\mu_{B}B^{z} $, and $J_0$ is the energy
difference between $S$ and $T_0$ states, shown in Fig.~\ref{fig:triplets}(b).
As the detuning $\varepsilon$ depends on time, the effective Hamiltonian
is also time-dependent.

In order to numerically model the experiment of Petta et al.\cite{petta10}
and compare the obtained singlet probabilities with the
experimental data, we write the wave function of the system
in the basis of the four singlet
and triplet states, and the singlet and triplet probabilities
are calculated from the wave function \cite{sarkka:045323}. The dynamics
of the wave function is calculated using the
relation $\psi(t+\Delta t)=\exp(-i\Delta t H_{eff}(t)/\hbar)\psi(t)$.

In our simulation, we describe the voltage by a detuning parameter $\varepsilon$
which has length as its unit. 
For the initialization at $t$=0, we set detuning
3 nm above its minimum value, which guarantees that the system is
initialized in the singlet state. 
We model the sweeping of the detuning by 
lowering it to a minimum value in 0.2 ns. 
This minimum value varies
between 2.5 and 5.5 nm.

\begin{figure}
\mbox{}\hfill\includegraphics[width=0.45\columnwidth]{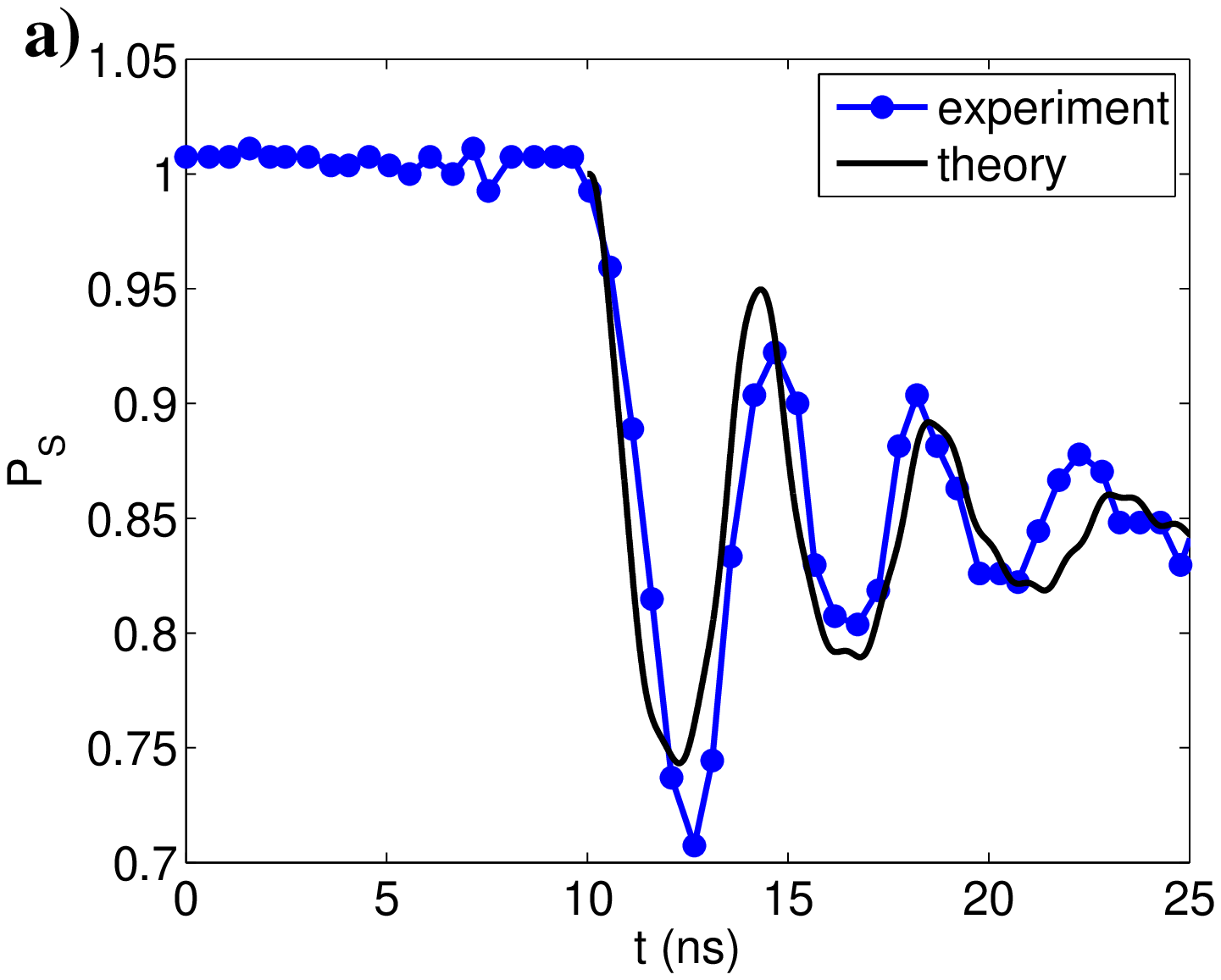}
\includegraphics[width=0.45\columnwidth]{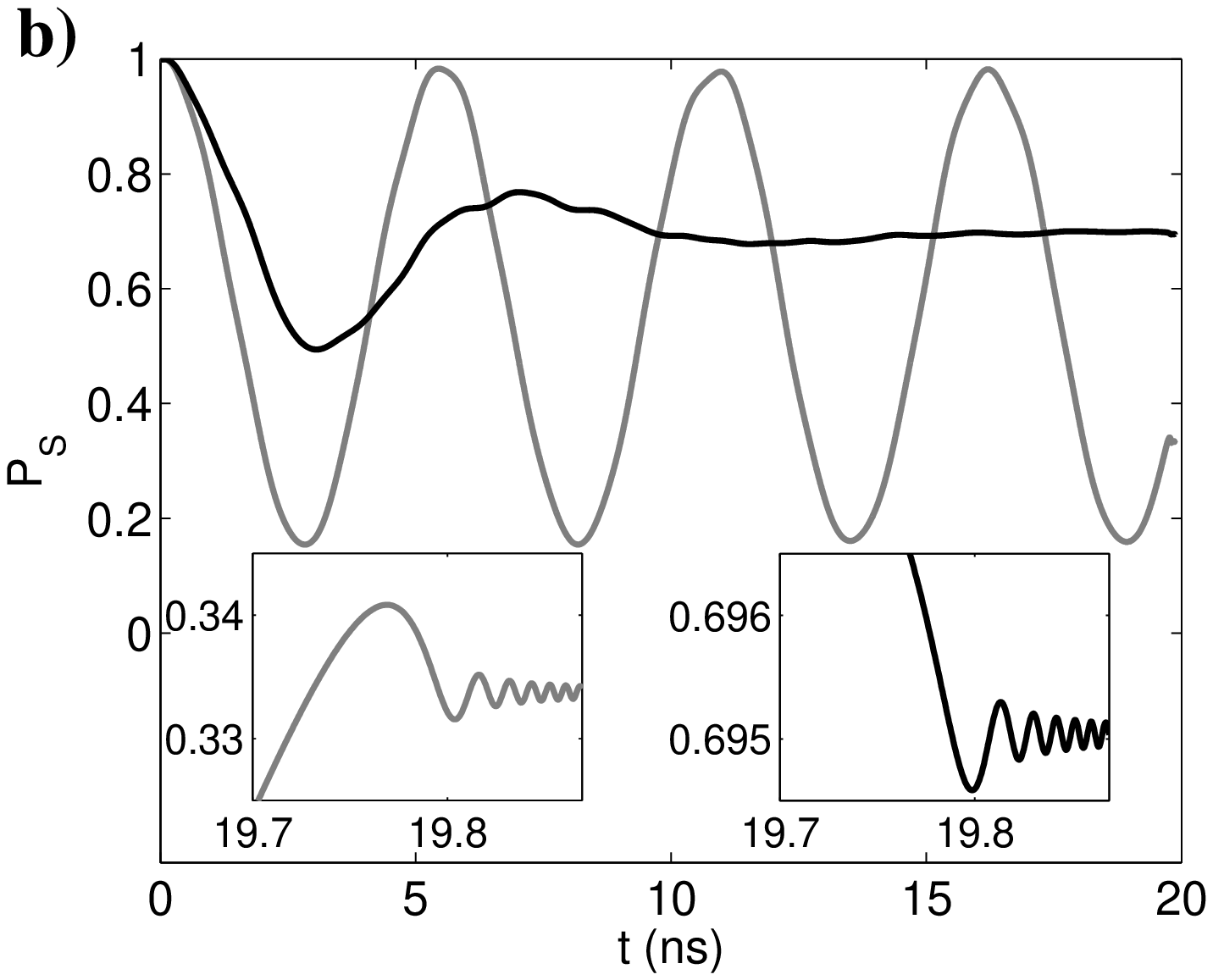}
\caption{(Color online)
\label{fig:sweep}
(a)
The singlet probability $P_{S}$,
as a function of the time $t$.
The blue spheres correspond to measurement results
of Petta et al. (upper part of Fig.~3(d) in
Ref.~\cite{petta10}), black curve represents our theoretical
fit for detuning $\varepsilon$=4.0 nm, averaged
over 1000 realizations. The theoretical data is shifted by 10 ns, 
because there is a delay in the experiments before singlet-triplet oscillations emerge.
(b)
The singlet probabilities,
as a function of the time $t$. The detuning $\varepsilon$
is 4.3 nm. The gray line is singlet probability calculated for a
single realization, black line is an average over 1000 realizations.
The insets show the probabilities when detuning is changed back to its
initial value in 0.2 ns. After passing the Landau-Zener crossing
during the detuning sweeping back to its initial value at 19.8 ns,
the oscillations are rapidly damped.}
\end{figure}

The parameters used in our simulation are determined by fitting
the theoretical results to the experimental
singlet probability shown in the upper part of Fig.~3(d) of
Ref.~\cite{petta10}. The fitting is shown in
Fig.~\ref{fig:sweep}(a). 
We set the detuning in our fit to $\varepsilon$=4.0 nm so that the 
oscillation periods,
which depend only on the $S$-$T_+$ energy difference, are the same.
We omit the first 10 ns of the experimental data when the
singlet probability stays constant. After 10 ns,
the experimental singlet probability drops to 0.7 and
then oscillates with period around 4 ns. 
The oscillation decays towards value 0.84. 
We set the hyperfine field strength and the sweeping time
to such values that the theoretical singlet-triplet oscillation
approaches 0.84 and the oscillation decays similarly with
the experimental data. Best fit is obtained with 
uniformly distributed hyperfine field values between
-7.5 mT and 7.5 mT. The root mean square of the hyperfine field
is 4.3 mT. In the experiments, the rms value has been around 2 mT \cite{taylor:035315}. 
The difference between these values might be due to the
different strengths of the confinement potential used in the
experiments and numerical model.
The resulting hyperfine coupling 
of the singlet and triplet states $\delta h^z$ is tens of neV, 
which is in the same magnitude than the experimental value 60 neV. 
We average the spin dynamics over 1000 random hyperfine
field realizations. As the sum of uniformly distributed
random numbers is Gaussian, our simulation corresponds
to a Gaussian distributed hyperfine field.
We sweep the detuning to the minimum value
in 0.2 ns. With these parameter values, the
decaying oscillations are similar with the experimental data.
Although the measured value may deviate from real
singlet probability, this fit gives the relevant scale for the
parameters.

The decay of the oscillations is due to the finite temperature
of the nuclear spins. Figure ~\ref{fig:sweep}(b) shows singlet
oscillations calculated for a single realization and an average
over 1000 realizations. We observe for the single realization a 
non-decaying oscillation.
For different realizations, the oscillations have different periods
and amplitudes. As a result, the averaged oscillations decay
and approach a constant value, as the solid line in Fig.~\ref{fig:sweep}(b) 
indicates.
The detuning is kept in the minimum value for 0-28 ns. The sweeping
of the detuning back to the initial value in 0.2 ns does not
significantly change the singlet probability. If we would model
the experiment rigorously, we should calculate separately each detuning cycle
for different waiting times, sweep the detuning back to the original value, and
save the singlet probability at the end. However, the singlet probability
does not significantly change during the return sweep. In the insets of
Fig.~\ref{fig:sweep}(b), the singlet-triplet oscillations during the 
sweeping back to initial detuning are shown for a single realization and for an average
over 1000 realizations.
The sweeping takes place between 19.7 ns 
and 19.9 ns. When Landau-Zener crossing point
is passed at 19.8 ns, the singlet-triplet oscillation stops and after that the faster
oscillations decay very rapidly. In practice, the oscillation freezes at
the value it had at the crossing point. During the 0.1 ns sweep from the
minimum value of the detuning to the crossing point, the singlet probability
changes less than 0.01. Hence, we may end the simulation
just before the detuning sweep and approximate the final value
of the singlet probability by using the value the probability has
just at the beginning of the detuning sweep.
This approximation enables calculation of the final
singlet probabilities for each waiting time in a single run.
We only have to make new runs for different detuning values.
We set the magnetic field to 100 mT, as in the experiment.

The analysis of the experiment of Petta et al. could be done using a three-level model, 
as it turns out that the $T_-$ state remains unoccupied. It might even be possible to
obtain an exact analytical solution for the singlet 
probability. A similar model was used to study a recent
experiment \cite{sun:51}, where a Josephson qubit was coupled to two two-level systems.
In this case, the two-level systems are not coupled to each other, making the
calculation of the time dependence easier.
We have the additional difficulty of $T_+$-$T_0$ coupling, and the resulting
singlet probability should be averaged over Gaussian distribution. As our numerical
method is reasonably fast to calculate, an analytical approach would not be 
practical for our study.

\section{Results and Discussion}

\begin{figure}
\mbox{}\hfill\includegraphics[width=0.9\columnwidth]{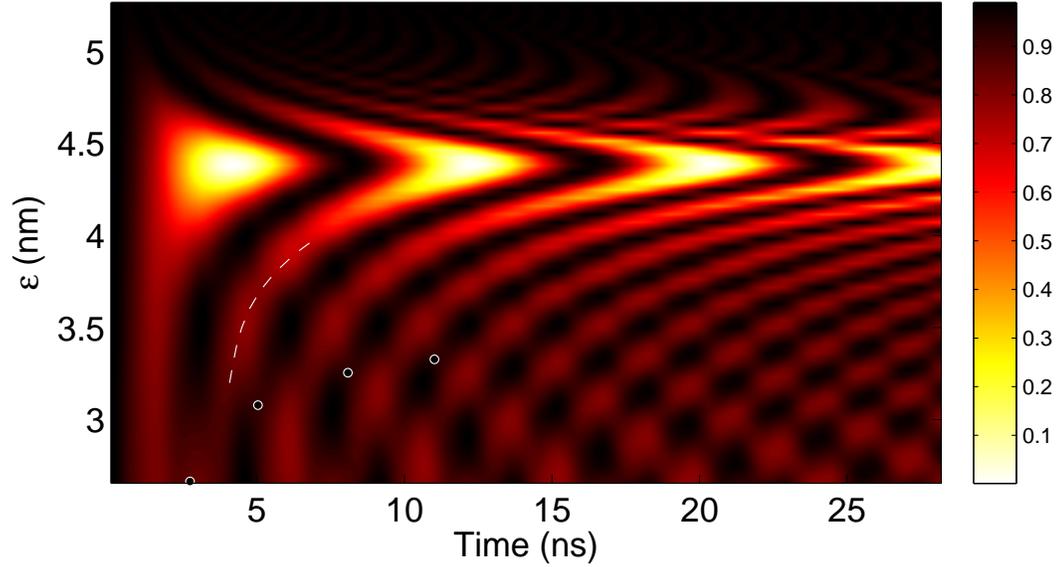}
\caption{(Color online)
\label{fig:fig3dd}
Probability $P_S$ of the
singlet state $S$ as a function
of time $t$ and detuning $\varepsilon$
calculated for a single hyperfine field realization.
The darker colors indicate increasing $P_S$.
The white dashed line indicates one of the interference lines
line due to the $S$-$T_+$ oscillation and black spheres
show the point in the interference lines where the
singlet probability is diminished due to the $S$-$T_0$
oscillation.} 
\end{figure}

\begin{figure}
\mbox{}\hfill \includegraphics[width=0.9\columnwidth]{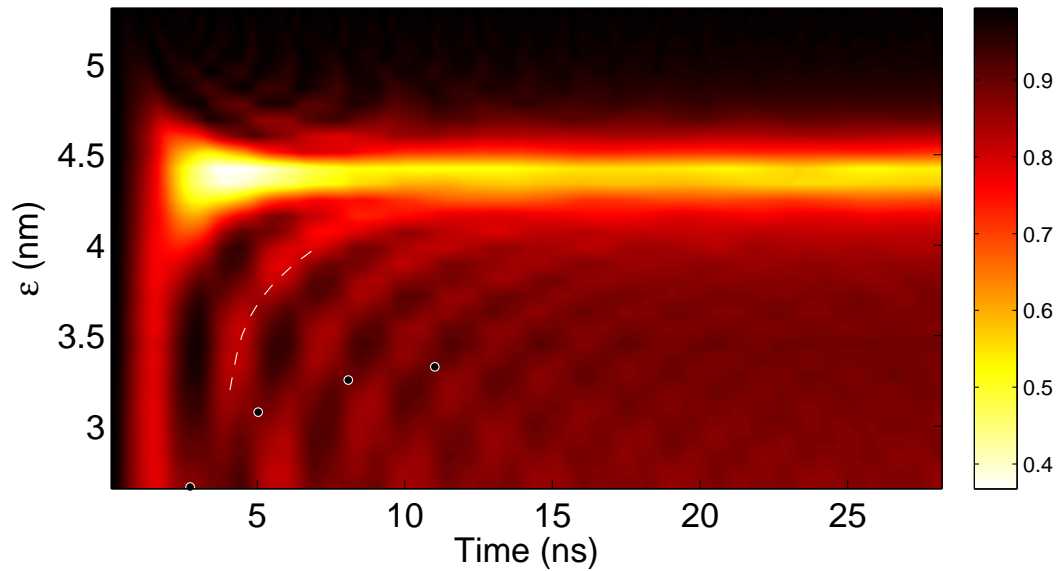}
\caption{(Color online)
\label{fig:fig3cc}
Same as Fig.~\ref{fig:fig3dd}, but
here the singlet 
probability $P_S$
is calculated using an average of 1000 hyperfine
field realizations and the color scale is different.} 
\end{figure}

Now after we have obtained the necessary parameters,
we can vary the detuning $\varepsilon$ and also compare
results with the experiment of Petta et al.\cite{petta10}.
The calculated singlet probabilities for detuning 
values between 2.5 and 5.5 nm are shown in
Fig.~\ref{fig:fig3dd} (single realization) and 
Fig.~\ref{fig:fig3cc} (average over 1000 realizations).
The single realization case is shown because the interference
effects are then easier to observe.


We have plotted the singlet
probability as a function of time (x-axis) and 
detuning (y-axis). The darker colors indicate increasing
singlet probability. For detunings above 
the $S$-$T_+$ crossing point $\varepsilon$=4.4 nm,
the singlet probability does not change considerably from
value 1, making the upper part of the figure dark.
The period of the singlet probability oscillations depends inversely
on the energy difference between $S$ and $T$ states.
In Fig.~\ref{fig:fig3dd}, we observe that the oscillation period
diminishes, when detuning values are further from the Landau-Zener crossing
and the $S$-$T$ energy difference increases.
At the crossing, the amplitude and period of the oscillation are largest.
The minima of the oscillations form observable red lines on
the dark background. One of them is accentuated with a 
white dashed line.
The same phenomena are present in Fig.~\ref{fig:fig3cc}, but the 
averaging makes the figure more blurred. 
When the detuning is near the crossing point, the singlet
oscillation drops from 1 to 0.5 and the later oscillations
are rapidly damped. This causes the yellow band
in the middle of the figure. 
The oscillations for detunings
above the crossing point are barely visible. 
For detunings under the
crossing point, the oscillations are damped and the bands formed by
the minima of the probability are not so clear.

\begin{figure}
\mbox{}\hfill \includegraphics[width=0.8\columnwidth]{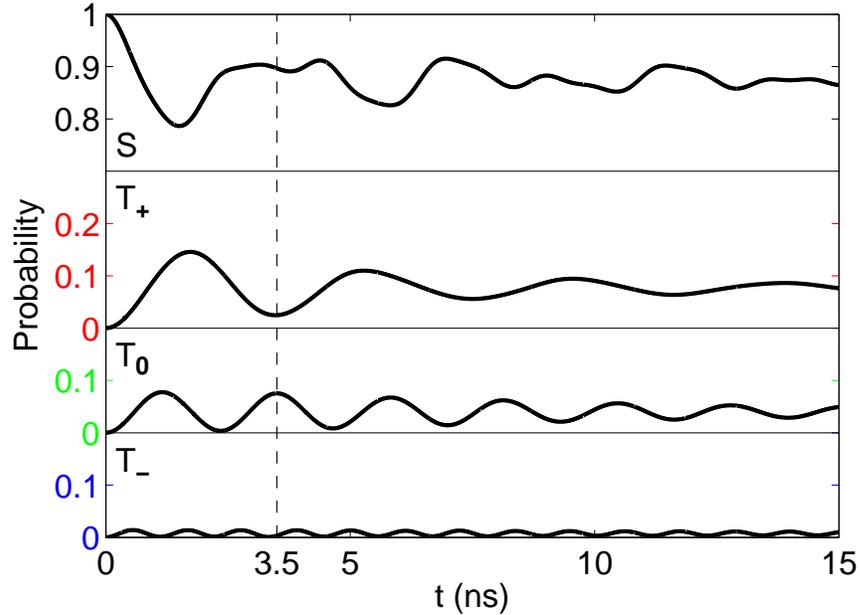}
\caption{(Color online)
\label{fig:interf}
The singlet and triplet probabilities,
as a function of the time $t$. The detuning $\varepsilon$
is 2.9 nm. The dashed line at 3.5 ns marks the point
where the occupation of $T_0$ state
diminishes the probability of the $S$ state.}
\end{figure}

The occupation probability of the $T_0$ state
affects the singlet-triplet oscillations by damping the oscillations
for certain detuning values. The red interference lines are then
slightly narrower. These narrower places in the lines form
a secondary interference pattern, one of those is indicated with black
spheres in the figure. This pattern is better seen in Fig.~\ref{fig:fig3dd},
where the widening and narrowing of the lines is clearly observable.
For example, for detuning 
$\varepsilon$= 2.9 nm, the interference of the 
triplet states at 3.5 ns causes a dip in the peak
of the singlet probability in Fig.~\ref{fig:interf}.
This effect is quite small and not easy to observe
experimentally.
In the situation where the energy of the 
$S$ state is closer in energy to $T_0$ state
than to $T_+$ state,
the period is given by the $S$-$T_0$ energy difference. Hence,
when these two energy differences are equal at $\varepsilon$=3.3 nm,
the singlet-triplet oscillation period has its smallest value 3.2 ns.
The interference lines are not symmetric with respect to
the detuning of the singlet-triplet crossing point.
This is due to the asymmetry of the exchange energy around
the crossing point as a function of detuning. As the exchange energy 
increases rapidly
for detunings above the crossing point, the period of the
oscillation decreases swiftly and the interference lines
disappear. The features in the figure above $\varepsilon$=4.5 nm
would vanish completely if the slope of the exchange energy were
steeper.

When we compare our results with experiments of
Petta et al. \cite{petta10}, we find that the interference
patterns of the oscillations are quite similar.
The contrast of the interference lines is better in the
experimental figure. This might be due to the small number
of detuning sweeps used in the averaging of due to correlations
between different sweeps.
The experimental pattern
has an area where the singlet probability remains 1 before
the oscillation begins. The oscillations start later
for higher values of detuning, probably due to the smoothed voltage
pulse profile used in the experiment. We did not model this effect
in our calculation.
For detunings above the $S$-$T_+$ crossing, our data shows small oscillations
that are missing from the experimental data. These are, however, 
much smaller than in the model of Ref.~\cite{ribeiro:115445}.
Probably the measurement accuracy is not sufficiently high for
the observation of these small oscillations.
At the singlet-triplet crossing ($\varepsilon \approx$ 4.4 nm),
the numerical singlet probability is smaller than 0.5
at 5 ns. In the experiment this minimum is not seen,
but the singlet probability goes to 0.5 without having
decaying oscillations. This might be due to the smoothed
pulse profile, which causes over 15 ns delay of the 
singlet-triplet oscillation at the $S$-$T_+$ crossing.

In the measurements of Petta et al. \cite{petta10}, the 
interference effects due to the occupation of $T_0$ state are not visible. 
The resolution of the experimental data is not high enough
to observe the interference clearly, but if we compare the data with 
Fig.~\ref{fig:fig3dd}, we find that the widening and narrowing
of the bands could have taken place in the experiments. With the
experimental resolution, the oscillations for the lower detunings
look quite similar.
If the singlet probability could be measured more accurately, it would
be possible to detect this interference and observe whether
both $T_0$ and $T_+$ states are occupied.
If it is the case that the experiments do not exhibit this
interference, the reason for the disappearance of $S$-$T_0$ oscillations could
be that the $S$-$T_0$ coupling is much stronger than $S$-$T_+$ coupling, 
and the oscillations decay very rapidly. Or the $S$-$T_0$ coupling is much 
weaker and the $S$-$T_0$ oscillation period is longer than the measurement time.

A small nuclear polarization ($\sim$0.5\%) could prolong the
singlet-triplet dephasing time by two orders of magnitude, as predicted 
by Ramon and Hu \cite{ramon:161301}.
Then, the polarization would not change the $S$-$T_+$ crossing point but could weaken the 
$S$-$T_0$ coupling so that the $T_0$ state does not become occupied during the measurement,
even for detunings in the vicinity of $S$-$T_0$ crossing.
In that case, the interference effects due to $T_0$ state would vanish and
the smallest possible period of singlet-triplet oscillations would be given by $S$-$T_+$
oscillations at the Zeeman energy, resulting in a minimum period 1.6 ns observed by 
Petta et al. \cite{petta10}.

%The increase in the hyperfine field gradient
%was demonstrated by Foletti et al. \cite {foletti:903}. 
%when they tuned the $S$-$T_0$ decoherence by a nuclear spin pumping cycle.
%However, Reilly et al. \cite {reilly08} claim that they decreased the 
%gradient by similar pumping cycle, but the interpretation of their
%results has been questioned \cite{bluhm:216803}.
%The effect of measurement on the hyperfine field and the effects due to the dynamical nuclear 
%polarization need more analysis, before conclusive interpretations can be made on the 
%measurements.

The results we obtained do not considerably depend on the exact form
of the potential. We used the potential given by Eq.~(\ref{eq:pot}),
but any other smooth double-well potential would produce similar results
for the singlet-triplet oscillations.
The time dependence of the detuning was taken to be linear here. Other 
monotonic time dependencies
would not have great effect on the singlet probabilities, because the
transitions from singlet to triplet states occur at the crossing points,
where the time dependence can be approximated by a linear fit. 
The form of the time dependence does not matter outside the crossing points.

In summary, we have studied singlet-triplet
decoherence in a two-electron double quantum dot with an
external detuning voltage.
We calculated the singlet probability
as a function of time for different detunings.
The interference pattern of the singlet probability
is very similar with the recent experimental results
and the decay of the oscillations is described 
somewhat better than in another theoretical study of 
the same experiment \cite{ribeiro:115445}.
Our results indicate that also
$T_0$ state could be occupied, even if $S$-$T_0$
crossing point is not within the range of
detuning values used. When the system is in the
proximity of the crossing point, transition to
$T_0$ state is possible. 
In addition, we predict
an observable effect in the interference pattern
due to the occupation of $T_0$ state. 
If the measurement
process affects the $S$-$T_0$ coupling but not the $S$-$T_+$ coupling,
the interference effects could disappear, which we suggest
might be the case in the measurement of Petta et al.\cite{petta10}.

\section*{Acknowledgments}
\addcontentsline{toc}{section}{Acknowledgments}

We acknowledge the support of Academy of Finland through 
its Centers of Excellence Program (2006-2011).

\section*{References}
\addcontentsline{toc}{section}{References}

\bibliography{spindyresub}
\bibliographystyle{iopart-num}
\end{document}